\documentclass[conference]{IEEEtran}

\usepackage{cite}
\usepackage{amsmath,amssymb,amsfonts}
\usepackage{algorithmic}
\usepackage{graphicx}
\usepackage{textcomp}
\usepackage{xcolor}

\usepackage[inline]{enumitem}
\newcommand{\step}[2][]{\raisebox{.5pt}{\normalfont\textcircled{\raisebox{-.5pt}{\footnotesize#2\raisebox{0.5pt}{\footnotesize #1}}}}}

\newcommand{\smallstep}[2][]{\raisebox{.5pt}{\normalfont\textcircled{\raisebox{-.5pt}{\fontsize{7}{8}\selectfont#2\raisebox{0.5pt}{\footnotesize #1}}}}}

\begin{document}

\makeatletter
\def\ps@IEEEtitlepagestyle{%
\def\@oddfoot{\parbox{\textwidth}{\footnotesize
Author's version of a paper accepted for publication in Proceedings of the 2024 8th Cyber Security in Networking Conference (CSNet). 
\\
\textcopyright{} 2024 IEEE. 
Personal use of this material is permitted.  
Permission from IEEE must be obtained for all other uses, in any current or future media, including reprinting/republishing this material for advertising or promotional purposes, creating new collective works, for resale or redistribution to servers or lists, or reuse of any copyrighted component of this work in other works.\vspace{1.2em}}
}%
}
\makeatother

\title{Adaptive Optimization of TLS Overhead for Wireless Communication in Critical Infrastructure}

\author{\IEEEauthorblockN{%
Jörn Bodenhausen\IEEEauthorrefmark{1}, 
Laurenz Grote\IEEEauthorrefmark{1}, 
Michael Rademacher\IEEEauthorrefmark{3}\IEEEauthorrefmark{4},
Martin Henze\IEEEauthorrefmark{1}\IEEEauthorrefmark{4}
}
\IEEEauthorblockA{%
\IEEEauthorrefmark{1}\textit{Security and Privacy in Industrial Cooperation}, RWTH Aachen University, Germany \\
\IEEEauthorrefmark{3}University of Applied Sciences Bonn-Rhein-Sieg, Germany\\
\IEEEauthorrefmark{4}\textit{Cyber Analysis \& Defense}, Fraunhofer FKIE, Germany\\
\{bodenhausen, henze\}@spice.rwth-aachen.de \(\cdot\) laurenz.grote@rwth-aachen.de \(\cdot\) michael.rademacher@h-brs.de
}
}

\maketitle

\begin{abstract}
With critical infrastructure increasingly relying on wireless communication, using end-to-end security such as TLS becomes imperative.
However, TLS introduces significant overhead for resource-constrained devices and networks prevalent in critical infrastructure.
In this paper, we propose to leverage the degrees of freedom in configuring TLS to dynamically adapt algorithms, parameters, and other settings to best meet the currently occurring resource and security constraints in a wireless communication scenario.
Consequently, we can make the best use of scarce resources to provide tightened security for wireless networks in critical infrastructure.
\end{abstract}

\begin{IEEEkeywords}
Transport Layer Security, Wireless Networks, Critical Infrastructure.
\end{IEEEkeywords}

\section{Motivation}

Modern critical infrastructure is increasingly interconnected, while simultaneously being deployed over significantly larger areas~\cite{el2017analysis}.
Examples of this trend range from wind parks over power grids to smart cities.
However, the special characteristics of such widespread environments often render wired communication infeasible and thus call for the use of cost-efficient wireless network technology such as the 450 MHz LTE-M network for critical infrastructure in Germany~\cite{bodenhausen2023challenges}. %

\begin{figure}[t]
  \centering
  \includegraphics[width=\linewidth]{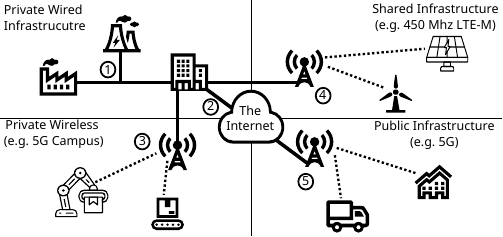}
  \caption{With critical infrastructure becoming more widespread and interconnected, a shift from traditional wired networks  \smallstep{1}, \smallstep{2} to wireless networks with private \smallstep{3}, shared \smallstep{4}, and public \smallstep{5} infrastructure becomes necessary.}
  \label{fig:infrastructure}
\end{figure}

However, the shift to wireless communication as visualized in Figure \ref{fig:infrastructure} comes with severe security implications and challenges. %
Even though the assumption of security through physical separation of a network \step{1} has long been invalid due to their connection to the Internet \step{2}, deployments are still not adequately protected \cite{dahlmanns2022missed,krause2021cybersecurity}.
For wireless technology, the implications are even more profound, as security goals are particularly easy to compromise even for private infrastructure \step{3} \cite{osanaiye2018denial,michaelides2024secure}.
Furthermore, wireless network infrastructure is commonly provided by a cellular operator and thus shared with other entities in a dedicated network \step{4} or even entirely public and routed via the Internet \step{5}.
Thus, any communication potentially traverses third-party infrastructures. 

The most promising approach to address resulting security concerns is end-to-end security, even if other security mechanisms are in place~\cite{henze2014trustpoint}.
In fact, regulators often demand the use of TLS, the most prominent end-to-end security approach, for communication in critical infrastructure, especially when using wireless communication \cite{bodenhausen2023challenges, rademacher2022bounds}. 
However, besides all advantages such as flexibility and interoperability, the use of TLS can constitute significant overhead for resource-constrained devices and networks \cite{rademacher2022bounds}.
Still, and providing the main motivation for this work, this overhead is not static as it depends on concrete parameterization, opening the potential to optimize the TLS overhead for specific scenarios.

\textbf{Related Work.}
Various works study the overhead of TLS through measurements, e.g., with regard to
\begin{enumerate*}[label=(\roman*)]
\item constrained LoRaWAN networks  \cite{rademacher2022bounds}, 
\item energy \cite{albela2018practical}, or
\item CPU, memory, and bandwidth overhead in TLS-secured MQTT \cite{dimov2022resource}.
\end{enumerate*}
While these works only focus on a particular setting or limited set of parameters, they still indicate that a trade-off for particular resources is possible.
For example, Restuccia et al.~\cite{restuccia2020low} identify memory overhead variations between TLS and DTLS as well as different implementations.
Moreover, for post-quantum algorithms, depending on the used network technology, either bandwidth or computational time is the main limiting factor for TLS connection establishment \cite{gonzalez2022kemtls}.
Furthermore, proposed optimization efforts are generally focused on particular settings.
For instance, Lauer et al. \cite{lauer2020analysis} focus on optimizing cryptographic computations on hardware-accelerated devices.

\textbf{Contributions.} 
To adaptively optimize TLS for wireless communication in critical infrastructure and thus enable its widespread use even in challenging scenarios, we propose a two-step approach.
First, to obtain a thorough understanding of the TLS overhead, we perform comprehensive measurements along several dimensions, covering all potentially practically relevant settings and a multitude of algorithms and parameters (Sec.~\ref{sec:overhead}).
Second, we turn these insights into use by designing and implementing an approach that can dynamically choose and adjust TLS parameters to meet resource constraints of a given scenario and adapt to changes, e.g., in available bandwidth (Sec.~\ref{sec:optimization}).
Moreover, to illustrate how this approach can be utilized and implemented in real-world deployments, we describe its application to a particular use-case scenario and demonstrate the significance of TLS bandwidth optimization in this scenario through practical measurements (Sec.~\ref{sec:usecase}).

\section{Variability in the Overhead of TLS} \label{sec:overhead}

Any efforts towards optimization of TLS require a profound understanding of the resulting overhead and its variability under certain conditions. 
While related work already provides valuable insights into isolated aspects of TLS overhead, a comprehensive picture of all influence factors across all relevant dimensions of TLS overhead is still missing.
To fill this gap, we propose a comprehensive measurement setup to study variability in TLS overhead (Sec.~\ref{sec:overhead:design}) and then exemplary showcase the potential for optimizing the bandwidth consumption of TLS especially in wireless networks (Sec.~\ref{sec:overhead:results}).

\subsection{Comprehensive Measurement Setup} \label{sec:overhead:design}

\begin{figure}[t]
  \centering
  \includegraphics[width=\linewidth]{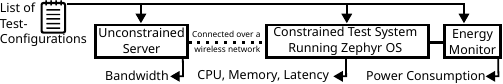}
  \caption{Our comprehensive measurement setup enables the analysis of the overhead of TLS across various dimensions.}
  \label{fig:setup}
\end{figure}

To lay the foundation to thoroughly study the impact of various TLS configurations, we outline the design of a comprehensive measurement setup to evaluate the TLS overhead across various dimensions.
Our intended setup is illustrated in Figure \ref{fig:setup} and consists of an unconstrained server, a constrained system that shall be evaluated, as well as an energy monitor. 
Initially, the devices are provided with a list of TLS configurations that shall be tested. 
The constrained device then iterates over the list and connects to the server with the respective configuration while the overhead is measured on the respective devices. 
Bandwidth is evaluated on the unconstrained server, as the client might not be capable of creating network captures. 
CPU, memory, and latency overhead must be evaluated directly on the constrained device. 
For the evaluation of power consumption, additional hardware is required similar to the setup from Suárez-Albela et al.~\cite{albela2018practical}.
While we intend to evaluate each overhead dimension independently to avoid interference, our proposed setup promises to provide valuable insights into their interdependency and to identify trade-off potential.

\subsection{Case Study: Bandwidth Overhead} \label{sec:overhead:results}

To better illustrate the potential of the proposed comprehensive measurement setup, we will exemplary showcase the huge variability in TLS bandwidth overhead depending on the TLS configuration. 
More specifically, we study the bandwidth required for different TLS versions and authentication mechanisms as well as varying elliptic curves.
To this end, we utilize a custom measurement setup that can capture local TLS communication generated by OpenSSL 3.2.1, wolfSSL 5.6.6, or Mbed TLS 3.6.0 respectively. 
For each parameter combination, we perform 30 runs and the observed variance can be traced back to varying acknowledgements and negligible variance in the generated keys and signatures.
We perform a full handshake with mutual authentication via a single self-signed X.509 certificate as generated by OpenSSL and subsequently exchange two 128 byte messages.
In the following, we report on the arithmetic mean over the runs of each setting and their standard deviation.

\begin{figure}[t]
  \centering
  \includegraphics[width=\linewidth]{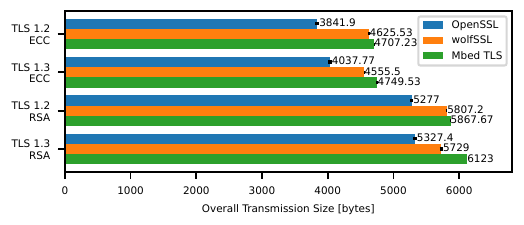}
  \caption{The average bandwidth overhead of a full TLS handshake varies widely across different authentication mechanisms as well as TLS libraries.}
  \label{fig:bandwidth}
\end{figure}

\begin{figure}[t]
  \centering
  \includegraphics[width=\linewidth]{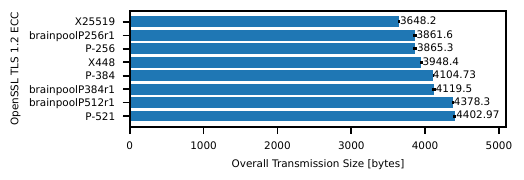}
  \caption{While the bandwidth overhead of TLS messages mainly depends on the security level of the used ECC curve, small variations exist even for curves with the same security.}
  \label{fig:curves}
\end{figure}

A first measurement illustrated in Figure \ref{fig:bandwidth} highlights the significance of the authentication mechanism on the overall bandwidth overhead. 
Moreover, a significant difference between libraries due to different default configurations and message processing can be observed.
Our second measurement illustrated in Figure \ref{fig:curves} considers various elliptic curves used within an \textit{ECDHE\_ECDSA} connection establishment. 
Since various TLS messages are influenced by this particular choice, elliptic curves are an excellent example of a particular configuration choice that yields a trade-off potential between bandwidth overhead, security level, and further dimensions not yet examined by us (e.g., energy \cite{albela2018practical}).

Our preliminary results provide several insights.
First, there is indeed a huge potential to optimize TLS overhead, especially for bandwidth-constrained communication.
This can already be seen from the varying bandwidth between default configurations of libraries for a fixed key-exchange method. 
Moreover, a trade-off between security and bandwidth is possible, as an increase in security due to larger key sizes directly results in additional bandwidth overhead.
Lastly, comprehensive measurements are required to uncover variances and differences between different implementations and scenarios.

\section{Adaptive TLS Overhead Optimization} \label{sec:optimization}

\begin{figure}[t]
  \centering
  \includegraphics[width=\linewidth]{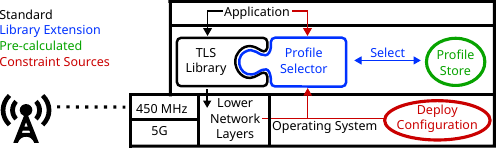}
  \caption{We propose to extend a TLS library with a profile selector that chooses algorithms and parameters tailored to current resource constraints based on pre-computed profiles.}
  \label{fig:concept}
\end{figure}

The overhead of TLS is not static, but depends on the particular configuration and parameters, allowing for a trade-off between different optimization dimensions \cite{restuccia2020low, gonzalez2022kemtls}.
To leverage this potential, we propose to automatically adapt TLS parameters to achieve the best trade-off for the current resource and security constraints.

Our basic concept for such adaptive TLS optimizations is illustrated in Figure \ref{fig:concept}. 
Without our proposed optimizations, an application running on a constrained platform has to use a TLS library with a configuration that is chosen and fixed at compile time (black), preventing dynamic adaptations to resource constraints.
Our main idea to extend this setting with the ability to dynamically adapt the TLS configuration to changing resource-constraints is to extend the TLS library with a \emph{profile selector} (blue).
Without introducing breaking changes to the interaction between applications and libraries (thus affording backwards compatibility), the profile selector applies the best-fitting \emph{profile} (Sec.\ \ref{sec:optimization:profiles}), i.e., pre-defined configuration of a TLS library, based on the currently given \emph{resource and security constraints} (Sec.\ \ref{sec:optimization:constraints}).

\subsection{Profiles of TLS Configurations} \label{sec:optimization:profiles}

Profiles provide a collection of algorithms, parameters, and settings that represent a configuration of the TLS library together with information on the resulting overhead across all relevant dimensions and the security level. 
In particular, profiles will be generated for and adjusted to a given device and network in the development phase and deployed onto the device. 
During run-time, whenever changes in constraints are observed, the profile selector will select the best-fitting profile based on the new set of constraints.
The resulting precise tailoring to a particular setting promises to enable near-optimal utilization of the trade-off and optimization potential of TLS configurations.
Moreover, due to the pre-calculation of profiles, the run-time overhead of our approach is minimized.

\subsection{Dynamic Resource and Security Constraints} \label{sec:optimization:constraints}

Various sources of constraints have to be considered, most of which are dynamic, e.g., as they depend on current system utilization or condition of the wireless link.
These dynamics are immense as they are directly influenced by the current state of the underlying network, which the application is usually not aware of.
For example, in wireless networks, a connection may use a high-order modulation when a certain signal strength is available \cite{rademacher2021path}.
In other situations, a bad reception can significantly decrease throughput and increase the packet error rate, rendering timely transmission of TLS handshakes infeasible \cite{rademacher2022bounds}.
Furthermore, traffic patterns in critical infrastructure can be highly dynamic due to random events such as natural disasters, updates, or cyberattacks~\cite{sen2022contextual}.
In such events, a huge amount of devices may initiate simultaneous transmission, causing a shortage of radio resources or even interference.

To account for such challenges, the current TLS profile needs to be adapted dynamically based on these changing constraints.
Depending on the type of constraint, such updates can either be provided by the application or the operating system (cf.\ Figure \ref{fig:concept}).
For example, the application can provide meta-data regarding the urgency of the transmission and its security requirements.
Likewise, the operating system can extract information about the current reception from the modem.
While, in general, a single device has no information about the current or future overall network load, it can partially detect an overload situation due to an increased packet error rate or different transport layer information (e.g., TCP) and may defer its transmission or switch to another TLS profile.

\section{Use-Case Example} \label{sec:usecase}

\begin{figure}[t]
  \centering
  \includegraphics[width=\linewidth]{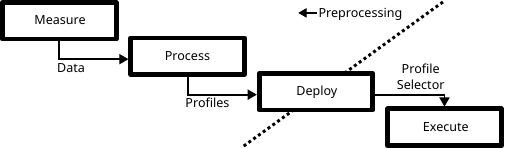}
  \caption{Much of the additional overhead, in particular the generation of profiles, would be performed in a pre-processing phase before the actual deployment, resulting in manageable processing overhead at run-time.}
  \label{fig:deployment}
\end{figure}

While our measurement setup (cf.\ Sec.\ \ref{sec:overhead}) promises to bring valuable and comprehensive insights into the interplay between TLS overhead in various dimensions and our profile selection mechanism (cf.\ Sec.\ \ref{sec:optimization}) aims to benefit from such insights, so far it remains open how this can be utilized in practice. 
Hence, in this section, we describe a use-case example to illustrate how those components can be integrated to result in real-world savings.
For this, we consider a critical infrastructure network that monitors and controls distributed energy systems wirelessly, e.g., by periodically reporting sensor measurements~\cite{eggert2014sensorcloud,henze2013maintaining}. 
We assume it to mainly utilize the 450 Mhz LTE-M network~\cite{bodenhausen2023challenges}, with 5G serving as a backup.
This results in potential switching between shared and public infrastructure (cf. Figure \ref{fig:infrastructure}), which could lead to varying security requirements and resource constraints due to technical limitations of these technologies. 
Moreover, available bandwidth for each technology might fluctuate depending on the situation (cf.\ Sec.\ \ref{sec:optimization:constraints}).

In such a setting, detailed measurements can yield interesting insights to improve efficiency at run-time.
For instance, detailed measurements of the reception of the 450 Mhz network throughout a building allow for insights into attenuation effects \cite{sorgatz2024measurements}.
At run-time, such information can be utilized to estimate the available bandwidth at a certain deployment location. 
The different steps required to make such information usable within our proposed profile selection mechanism is illustrated in Figure \ref{fig:deployment}. 
The foundation is formed by detailed measurements in the lab and in the wild for various TLS configurations. 
In a subsequent analysis phase, the results are examined in detail and insights about potential trade-offs are derived.
From this, a suitable set of profiles is generated once and integrated in the profile selector at compile time (though later updates are also possible). 
Thus, at run-time, merely a selection of suitable profiles needs to be performed by searching the profile store.
The data depicted in Figures \ref{fig:bandwidth} and \ref{fig:curves} already shows that a trade-off between security and bandwidth is possible, i.e., striving for the most secure key exchange mechanisms as long as the available bandwidth allows for it.

\begin{figure}[t]
  \centering
  \includegraphics[width=\linewidth]{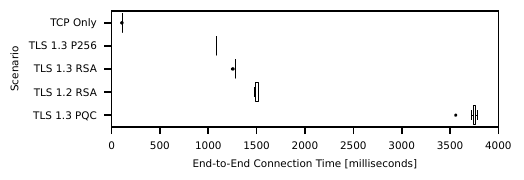}
  \caption{Introducing additional overhead due to large TLS handshakes into an LTE-M network leads to significant latency degradation.}
  \label{fig:amarisoft}
\end{figure}

To further illustrate the significance of the bandwidth overhead in particular on the overall performance of such a network, we performed a series of measurements.
The results of these measurements are shown in the boxplot in Figure \ref{fig:amarisoft}.
Here, we performed load tests on the 450 Mhz LTE-M Network between an Amarisoft Callbox and an Amarisoft UE Simbox simulating 64 devices.
We emulate polling of electrical substations with the IEC 60870-5-104 protocol, i.e., a 38 bytes request followed by a 282 bytes reply, in one second intervals.
Thus, the TLS handshake constitutes a significant overhead. 
We observe that the achievable end-to-end connection time in the network significantly degrades with the introduction of TLS and further degrades when utilizing post-quantum secure mechanisms. 
Here, e.g., an opportunistic use of post-quantum mechanisms would be conceivable with a fallback to conventional mechanisms in case of an overload situation to ensure availability.

While this illustrates that a trade-off generally is possible and can be utilized in a real-world scenario, focussing on security and bandwidth alone might not yield groundbreaking efficiency optimization on its own.
However, the possibility to benefit from further trade-off potential is extremely likely, as for instance highlighted by related work for computational overhead of certain post-quantum secure algorithms \cite{gonzalez2022kemtls} or energy consumption \cite{albela2018practical}, which is particularly relevant for battery-powered devices. 
Furthermore, such considerations for optimization can be extended to the timing of handshakes and the utilization of session resumption mechanisms.

\section{Outlook \& Conclusion} \label{sec:conclusion}

With our work, we strive to leverage the degrees of freedom in configuring TLS to dynamically adapt its configuration to best meet varying resource and security constraints, especially for wireless communication in critical infrastructure.
Currently, we are working on fully realizing the comprehensive measurements of variability in TLS overhead (Sec.~\ref{sec:overhead}) and implementing our concept of adaptive TLS overhead optimization (Sec.~\ref{sec:optimization}).
To this end, we are joining forces with a device manufacturer to integrate our solution into a commercial sensor platform for critical infrastructure that unites 450 MHz LTE-M and 5G communication capabilities, which will allow us to evaluate our approach in a real-world 450 MHz LTE-M network of a cooperating utility provider.
Already today, our case study results show huge potential for adaptive optimization of TLS overhead for wireless communication in critical infrastructure (Sec.~\ref{sec:usecase}).
With our concept for adaptive TLS overhead optimization, we outline an approach to leverage this potential with minimal changes to existing devices and applications, thus affording wide deployability. 

\section*{Acknowledgment}

Funded by the German Federal Office for Information Security (BSI) under project funding reference numbers 01MO23003B and 01MO23003D (PlusMoSmart).
The responsibility for the content of this publication lies with the authors.


\end{document}